\documentclass[aps, prl, twocolumn, groupedaddress, showpacs, floatfix,10pt]{revtex4-1}
\usepackage{amsmath}
\usepackage{amscd}
\usepackage{amsthm}
\usepackage{amssymb} \usepackage{latexsym}
\usepackage{euscript}
\usepackage{epsfig}
\usepackage{graphics}
\usepackage{array} 
\usepackage{enumerate}
 
\usepackage{pictexwd,dcpic}
\usepackage{graphicx, epstopdf, url, comment, overpic, rotating} %packages added by Tobi

\usepackage{color, xcolor}

\usepackage{hyperref}    

\newcommand{\rf}[1]{(\ref{#1})}

\newcommand{\R}{{\mathbb R}}
\newcommand{\N}{{\mathbb N}}

\def\txte{{\textnormal{e}}}

\newcommand{\cG}{{\mathcal G}}
\newcommand{\cH}{{\mathcal H}}
\newcommand{\cM}{{\mathcal M}}
\newcommand{\cZ}{{\mathcal Z}}

\begin{document}
\title{Coupled Hypergraph Maps and Chaotic Cluster Synchronization}

\author{Tobias B\"ohle$^{a}$, Christian Kuehn$^{a,b}$, Raffaella Mulas$^{c,d,e}$, and J\"urgen Jost$^{c,f}$}
\affiliation{
\mbox{$^{a}$Faculty of Mathematics, Technical University of Munich, Boltzmannstr.~3, 85748 Garching b.~M\"unchen, Germany}
\mbox{$^{b}$Complexity Science Hub Vienna, Josefst\"adter Str.~39, 1080 Vienna, Austria}
\mbox{$^{c}$Max Planck Institute for Mathematics in the Sciences, Inselstr.~22, 04103 Leipzig, Germany}
\mbox{$^{d}$The Alan Turing Institute, The British Library, London NW1 2DB, UK}
\mbox{$^{e}$University of Southampton, University Rd, Southampton SO17 1BJ, UK}
\mbox{$^{f}$Santa Fe Institute for the Sciences of Complexity, 1399 Hyde Park Road Santa Fe, New Mexico 87501, USA}
}

\begin{abstract}
\noindent Coupled map lattices (CMLs) are prototypical dynamical systems on networks/graphs. They exhibit complex patterns generated via the interplay of diffusive/Laplacian coupling and nonlinear reactions modelled by a single iterated map at each node; the maps are often taken as unimodal, e.g., logistic or tent maps. In this letter, we propose a class of higher-order coupled dynamical systems involving the hypergraph Laplacian, which we call coupled hypergraph maps (CHMs). By combining linearized (in-)stability analysis of synchronized states, hypergraph spectral theory, and numerical methods, we detect robust regions of chaotic cluster synchronization occurring in parameter space upon varying coupling strength and the main bifurcation parameter of the unimodal map. Furthermore, we find key differences between Laplacian and hypergraph Laplacian coupling and detect various other classes of periodic and quasi-periodic patterns. The results show the high complexity of coupled graph maps and indicate that they might be an excellent universal model class to understand the similarities and differences between dynamics on classical graphs and dynamics on hypergraphs.
\end{abstract}

\maketitle

\section{Introduction}

There are many processes, which can be modeled elegantly using network dynamical systems. For example, this includes epidemic spreading~\cite{KissMillerSimon}, opinion formation~\cite{Liggett}, 
neural networks~\cite{HoppensteadtIzhikevich}, synchronization~\cite{Kuramoto}, and game theory~\cite{EasleyKleinberg}, just to name a few. The standard modeling approach for network dynamics is that each vertex (or node) has one or more suitable state variables while edges (or links) provide the interaction between the vertices. Yet, in many applications, just considering a standard graph model is insufficient and it is necessary to consider higher-order interactions, which can be modelled via hypergraphs. For example, higher-order coupling is of crucial importance in neuroscience~\cite{GiustiGhristBassett}, where it can encode co-firing of neurons, the joint activation of several brain areas as well as more general functional relationships between neurons or brain regions. In epidemiology~\cite{BodoKatonaSimon}, higher-order interactions can model the different epidemiological interaction patterns for group dynamics. Similarly to epidemiology, contact processes in opinion formation~\cite{HorstmeyerKuehn} clearly benefit from the inclusion of group dynamics. In cell biology~\cite{KlamtHausTheis}, classical graph structures can also fall short to correctly represent the interactions between proteins, metabolites, or genes, where chemical mechanisms actually dictate that certain binding or catalysis patterns only form, when more than two biological components react. Historically, the need to model higher-order interactions was observed very early within ecology~\cite{Abrams,BillickCase}, where the interaction between more than two species is very common in predator-prey systems due to intertwining of various competitive, mutualistic, or parasitic effects. Hence, higher-order interactions have recently become a common physical modelling principle and we refer to~\cite{BATTISTON20201} for a detailed survey of the area, which displays a high level of recent activity~\cite{MulasKuehnJost,Synchronization,Carletti,Moreno,Salova} in the context of dynamics. In this work, we are interested in the effect that higher-order coupling between nodes can have on the interactions between coupled maps. Coupled map lattices (CMLs) are a very classical simplified/universal class, which are quite well-studied on graphs. Yet, to the best of your knowledge, intertwining the dynamical properties of CMLs in the context of hypergraphs has not been considered. Here we propose a very general extension to chemical hypergraphs, which includes classical hypergraphs as a special case.
%\red{[TODO:JJ/RM:Please explain here the physical/chemical origin of your hypergraph Laplacian together with citations.]}
While classical hypergraphs are made by vertices that are joined together by hyperedges, chemical hypergraphs have the additional structure that each vertex--hyperedge incidence is given either a plus sign, a minus sign, or both. They were introduced in \cite{JM} as a model for chemical reaction networks, in which vertices represent chemical elements and hyperedges represent chemical reactions. The signs allow to distinguish reactants, products and catalysts within a chemical reaction. Moreover, in \cite{JM}, also a normalized Laplace operator for chemical hypergraphs was introduced. Such a operator is a natural generalization of the normalized Laplacian for graphs that was introduced by Fan Chung \cite{Chung}, and its spectrum, as in the graph case, encodes many qualitative properties of the hypergraph to which it is associated. Conceptually, there are two possible approaches for constructing hypergraph Laplacians. The first approach considers a random walk or a diffusion between the vertices of a hypergraph and identifies a Laplacian as the generator of such a process. Thus, one looks at transition probabilities between vertices. Since these are pairwise relations, however, they can be encoded as the weights of some ordinary graph. The resulting Laplacian can then be represented as the Laplacian of that underlying \emph{effective} graph. Thus, we would be back to graph theory. The second approach, the one that we are adopting here, considers relations between sets of vertices, for instance between the inputs and the outputs of reactions. Mathematically, one can define a boundary operator representing such relations, and when endowing the vertex and the hyperedge set with suitable scalar products, construct the corresponding Laplacian. Such a Laplacian, in contrast to the first approach, cannot be reduced to a graph Laplacian, but rather genuinely represents and reflects the higher order relations encoded by the hypergraph. The essential point is that the operator compares the total input of an oriented hyperedge to its total output, but this still allows the individual contributions of the input nodes to vary, and the same holds  for the output nodes. When nodes are contained in several hyperedges, they then are subjected to the constraints coming from each of these hyperedges, but it turns out that for many hypergraphs, there is still enough freedom to make genuinely new dynamical phenomena possible.\\
In this work, we employ that hypergraph Laplacian to focus our study of the new class of coupled hypergraph maps (CHMs) on chaotic dynamics and synchronization as these two phenomena are cornerstones in the study of more classical CMLs. Our work is structured as follows: First, we recall and develop the main mathematical tools we need for classical CMLs. Then we introduce CHMs defined via iterated maps at each node and hypergraph Laplacian coupling between nodes. In the main dynamics part, we study synchronization and chaos for the paradigmatic example of hyperflowers as well as some further interesting hypergraphs. Last, we give a brief conclusion and an outlook to future work.

\section{Coupled Map Lattices}
\label{sec:CML}

To study dynamical systems induced by iterating coupled maps on networks is an established paradigm in nonlinear dynamics originally termed coupled map lattices (CMLs)~\cite{Kaneko}. Each node/vertex $i\in\{1,2,\ldots,d\}$ of a connected network/graph $\cG$ evolves according to a time-discrete map $f$. Typical examples are the logistic map $f(x)=\mu x(1-x)$ or the tent map $f(x)=\frac\mu2\min\{x,1-x\}$ each with parameter $\mu\in[0,4]$. The dynamics of the state $x_n(i)\in\R$ at node $i$ at time $n\in \N$ is defined via 
\begin{equation}\label{1}
x_{n+1}(i)=  f(x_n(i)) - \epsilon (\Delta_\cG f)(x_n(i)),
\end{equation}
where $\epsilon\in\R$ is a parameter controlling the diffusive coupling, and the normalized Laplacian $\Delta_\cG$ is defined as
\begin{equation}\label{2}
(\Delta_\cG u)(x(i)):=u(x(i))- \frac{1}{\deg i}\sum_{j\sim i} u(x(j)), 
\end{equation}
where $i\sim j$ when $i,j$ are connected by a link/edge, and we call them neighbors in that case, and $\deg i$ is the number of neighbors of $i$. Classically, one has considered lattices $\cG=L(d_1,d_2)$ ($d_{1,2}\in\N$) with $d_1d_2$ nodes, or complete graphs $\cG=K_d$ ($d\in \N$) on $d$ nodes; both classes already display very surprising phenomena. Subsequently, triggered by the rise of network science, it was discovered that new effects may arise in CMLs when the graph is not complete or a lattice~\cite{JJ1,LindGallasHerrmann,WangXu1}. A commonly encountered theme in all classes of CMLs is synchronization~\cite{PikovskyRosenblumKurths}. Dynamics is called synchronized if $x_n(i)=x_n(j)$ for all $i,j$ and all times $n\ge n_0$ for some $n_0\in \N$. Importantly, synchronized dynamics need not be constant in $n$, but could, for instance, show itself chaotic behavior. In such a case, one speaks of the \emph{(complete) synchronization of chaos}~\cite{Kaneko1984}; we refer to \cite{FujisakaYamada} for the discovery of the general effect of chaotic synchronization in coupled oscillators. An important object is the (complete) synchronization manifold $\cM:=\{x(1)=x(2)=\cdots=x(d)\}$, and one is interested in the transverse stability of $\cM$. Consider an orbit $\bar{\gamma}=\{\bar{x}_n\}_{n=1}^\infty$ of the given map $f$. If the CML is uncoupled ($\epsilon=0$), then the homogeneous solution $\gamma= \{x_n(j)=\bar{x}_n\}$ for all $j$ is synchronized and remains in $\cM$. Then one may linearize around $\gamma$, derive a variational equation, and link stability for $\epsilon>0$ to the Lyapunov exponent of $f$ and the eigenvalues of $\Delta_\cG$. As shown in \cite{JJ1}, (complete) synchronization on $\cM$ is transversally linearly stable for~\eqref{1} if 
\begin{equation}\label{3}
| \txte^{\mu_0} (1 - \epsilon \lambda_k) | < 1\quad \forall k\in \{2,\ldots,d\},
\end{equation}
where
\begin{equation*}
\mu_0 = \lim_{N \rightarrow \infty} {\frac{1}{N}} \sum_{n=0}^{N-1} \log |f'(\bar{x}_n)| 
\end{equation*}
is the Lyapunov exponent of $f$ and $\lambda_k$ are the nonzero eigenvalues of the Laplacian \rf{2}, which can be ordered as
\begin{equation*}
0=\lambda_1 < \lambda_2 \le \dots \le \lambda_{d}.
\end{equation*}
The eigenvalue $0$ is simple because we assumed that $\cG$ is connected. Obviously, it suffices to check \rf{3} for the eigenvalues $\lambda_2$ and $\lambda_d$. Thus, in favorable cases, there is a certain range of values of the coupling parameter $\epsilon$ for which \rf{3} is satisfied. Note that the eigenfunctions for the simple eigenvalue $\lambda_1=0$ are the constants corresponding to the tangential direction of the synchronization manifold. If we assume temporal instability with $\mu_0>0$, then \rf{3} is not satisfied for $\lambda_1=0$. Thus, when \rf{3} is satisfied for all other eigenvalues, the constants are the only unstable directions at a synchronized state, and this precisely means that $\cM$ is linearly transversally stable. The concept of chaotic synchronization can be combined with the idea of \emph{cluster synchronization}, i.e., different subsets/clusters of nodes synchronizing among themselves, but not across clusters~\cite{Kaneko1,BelykhBelykhMosekilde}.

\section{Coupled Hypergraph Maps}

A hypergraph consists of vertices connected by hyperedges, which can couple more than two vertices. For chemical hypergraphs~\cite{JM}, we also additionally label vertices of each hyperedge as inputs and outputs; note that these classes need not be disjoint. The (chemical) hypergraph Laplace operator is then given by
\begin{eqnarray}
\nonumber
&\Delta_\cH u(x(i)):=\\
\nonumber
&\frac{\sum_{h_{\text{in}}: i\text{ input}}\biggl(\sum_{i' 
\text{ input of }h_{\text{in}}}u(x(i'))-\sum_{j' \text{ output of }h_{\text{in}}}
u(x(j'))\biggr)}{\mathrm{hypdeg}\ i}+\\
&-\frac{\sum_{h_{\text{out}}: i\text{ output}}\biggl(\sum_{\hat{i} \text{ input of }
h_{\text{out}}}u(x(\hat{i}))-\sum_{\hat{j} \text{ output of }h_{\text{out}}}
u(x(\hat{j}))\biggr)}{ \mathrm{hypdeg}\ i}\nonumber
\end{eqnarray}
with
\begin{equation*}
\mathrm{hypdeg}\ i:=\sum_{h, i\in h} (|h|-1),
\end{equation*}
where $|h|$ is the number of vertices contained in the hyperedge $h$. Here, the first sum in the definition of $\Delta_\mathcal H$ runs over $\{h_{\text{in}}: i\text{ input}\}$, which is the set of all hyperedges $h_\text{in}$, in which the $i$th node is classified as an input node. For a given $h_\text{in}$, the set $\{i' \text{ input of }h_{\text{in}}\}$ consists of all nodes which are an input of $h_\text{in}$ and the other summations can be explained analogously.
%\red{[TODO:JJ/RM:Also here I would believe that additional physics and physical motivation explaining the definition via a physical example would be appreciated by the physics community.]}
The hypergraph Laplacian $\Delta_\cH$ is a natural generalization of the classical normalized Laplacian for graphs in \eqref{2}. Observe that, given a function $u$, the graph Laplacian in \eqref{2} gives the difference between the value of $u$ at $x(i)$ and the average of the values of $u$ at the $x(j)$'s, where the $j$'s are the neighbors of the node $i$. The hypergraph Laplacian has a similar, but more complex, interpretation. In fact, the contribution of $u(x(j))$ in $\Delta_\cH u(x(i))$ depends on how many hyperedges the nodes $i$ and $j$ have in common, as well as on the orientations that $i$ and $j$ have on these hyperedges. For example, if two nodes $i$ and $j$ are contained in exactly two common hyperedges $h_1$ and $h_2$, and they have the same orientation in $h_1$ while they have opposite orientations in $h_2$, then $u(x(j))$ does not appear in $\Delta_\cH u(x(i))$, because the terms that correspond to $h_1$ and $h_2$ cancel each other.
%\textcolor{red}{[RM: Here we can write something about the phyisical interpretation. I thought we could describe a flow but I was confused with the edge Laplacian.] }
In order to give a more practical interpretation, we equivalenty re-write the hypergraph Laplace operator as follows. Given a node $i$ and an hyperedge $h$, let
\begin{equation*}
    \textrm{o}(i,h):=\begin{cases}
    1 & \text{ if }i\in h \text{ is an input }\\
     -1 & \text{ if }i\in h \text{ is an output }\\
      0 & \text{ otherwise.}
    \end{cases}
\end{equation*}Then,
\begin{equation*}
    \Delta_\cH u(x(i))=\frac{1}{\mathrm{hypdeg}\ i}\biggl(\sum_{h:i\in h}\mathcal{F}(i,h)\biggr),
\end{equation*}where
\begin{equation*}
  \mathcal{F}(i,h):= \sum_{\substack{i'\in h:\\ \textrm{o}(i',h)=\textrm{o}(i,h)}}u(x(i'))-\sum_{\substack{j'\in h:\\ \textrm{o}(j',h)=-\textrm{o}(i,h)}}u(x(j')).
\end{equation*}Hence, if we see $u(x(k))$ as the amount of a given quantity at node $k$, and if a given node $i$ is an input for a hyperedge $h$, then $\mathcal{F}(i,h)$ is the difference between the total amount of that quantity at the inputs of $h$, and the total amount of that quantity at the outputs of $h$. The closer $\mathcal{F}(i,h)$ is to zero, the more %the  quantity is equally distributed among the inputs and the outputs of $h$. The closer $\Delta_\cH u(x(i))$ is to %zero, the more the quantity is equally distributed among inputs and outputs, when considering at the same time all %hyperedges in which $i$ is contained.\newline
 the total amount of input  balances the total amount of output of $h$. But the individual input nodes can contribute  quite differently, as only their sum enters into the balance, and the same is true for the individual output nodes. This is the source of new phenomena for dynamics on hypergraphs governed by the Laplacian compared to what we can see on ordinary graphs.\newline The definition of $\Delta_\cH$ is modified from~\cite{JM}, in order to ensure an appropriate normalization for our dynamics. The hypergraph Laplacian $\Delta_\cH$ again has real spectrum. Analogously to \rf{1}, we want to couple the dynamics on a hypergraph via $\Delta_\cH$ for a given map $f:[0,1]\rightarrow [0,1]$ at each node. The hypergraph Laplacian may fail to satisfy the maximum principle; this is the case when it has non-constant eigenfunctions for the eigenvalue $0$.
While it is easy to see that the iterated map \eqref{1} on graphs leaves the unit cube $[0,1]^d$ invariant if $\epsilon \in [0,1]$, the nonexistence of a maximum principle for the hypergraph Laplacian causes the unit cube $[0,1]^d$ not to be invariant anymore when directly replacing $\Delta_\mathcal G$ in \eqref{1} by $\Delta_\mathcal H$, even if $\epsilon \in [0,1]$. Therefore, we define a periodic triangular function
\begin{equation*}
\sigma(x):=\begin{cases}
    x-2k &\text{ if } x\in [2k,2k+1]\\
    2(k+1)-x &\text{ if } x\in [2k+1,2k+2]
\end{cases} 
\end{equation*}
for $k\in \mathbb{Z}$ and put
\begin{equation}\label{5}
x_{n+1}(i)=  \sigma \left(f(x_n(i)) - \epsilon (\Delta_\cH f)(x_n(i))\right),
\end{equation}
which makes the unit cube $[0,1]^d$ invariant under the dynamics and does not influence important properties of the dynamics, such as synchronization and chaotic behavior, that we will consider below.

\section{Hypergraph Dynamics}

\begin{figure}[h]
	\centering
	\begin{overpic}[width = .45\textwidth]{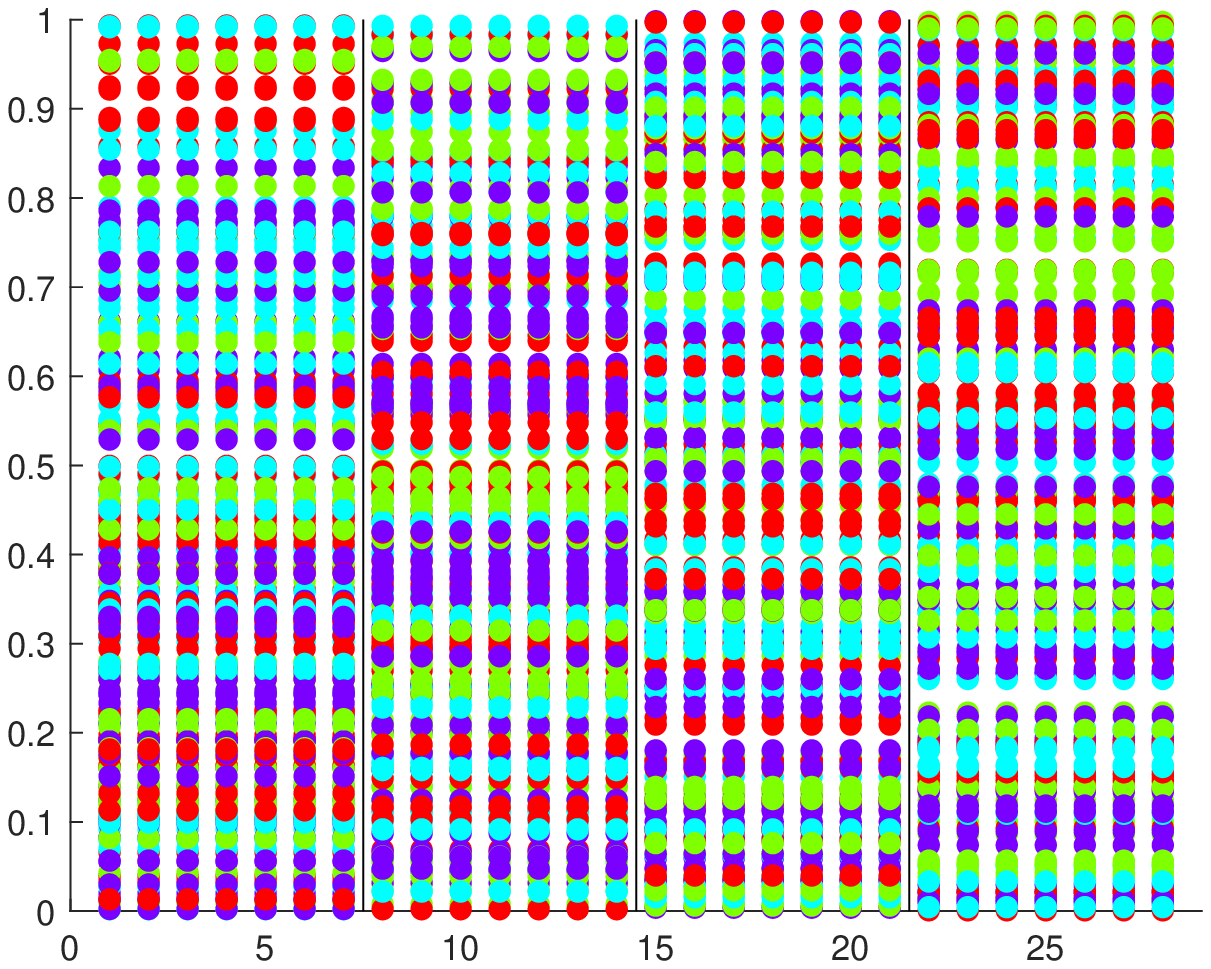}
		\put(3,35){\begin{turn}{90}$x_n(i)$\end{turn}}
		\put(18, 72){center}
		\put(36, 72){1st leaf}
		\put(54, 72){2nd leaf}
		\put(72, 72){3rd leaf}
		\put(48,1){Nodes $(i)$}
	\end{overpic}
	\caption{Numerical Integration of \eqref{5} for $f(x)=\mu x(1-x)$ on a hyperflower with $c=t=7$ and $\ell=3$ for $\mu = 1.4$ and $\epsilon = 8$. Plotted iterations are $5000< n \le 5200$. The values of $x_n(i)$ are alternately plotted in red, cyan, green and purple upon increasing $n$.}
	\label{fig:SingleDynamics}
\end{figure}

The spectral properties of $\Delta_\cH$ are richer than those of $\Delta_\cG$ \cite{JM}. In particular, $\Delta_\cH$ can possess the eigenvalue 0 with multiplicity $>1$, and none of the eigenfunctions need to be constants. An example is the hyperflower $\cH_{c,t,\ell}$ defined via three parameters $c$, $t$ and $\ell$ \cite{AndreottiMulas}. It is a generalization of the star graph. There is a set of $c$ \textsl{central vertices} and $\ell$ sets each consisting of $t$ \textsl{peripheral vertices}. Each set containing peripheral vertices is called a \textsl{leaf}. Central vertices are contained in all hyperedges, but each hyperedge additionally includes only peripheral nodes from one leaf, so in total there are $\ell$ hyperedges. By convention we classify central vertices as inputs and peripheral nodes as outputs. An example for the dynamics of~\eqref{5} in Figure~\ref{fig:SingleDynamics} shows complex chaotic cluster synchronization. 

To understand synchronization patterns, the results for CMLs on graphs motivate us to consider the eigenvalue/eigenfunction structure of hyperflowers $\cH_{c,t,\ell}$. The function which equals $-1$ on central nodes and $+1$ on peripheral nodes is an eigenfunction for the eigenvalue $(c+t)/(c+t-1)$. Next, we have functions that are $+1$ on one leaf, $-1$ one on another and $0$ elsewhere, corresponding to the second largest eigenvalue $t/(c+t-1)$. There are $\ell-1$ such linearly independent eigenfunctions. The remaining eigenfunctions have eigenvalue $0$. There is one eigenfunction, which attains the value $1/c$ on central nodes and $1/t$ on peripheral nodes. Furthermore, every function that is $+1$ on one node, $-1$ on another of the same component (the center or a leaf) and $0$ elsewhere is an eigenfunction; there are $c-1 +\ell(t-1)$ such linearly independent functions. Altogether, we have generated $c+t\ell$ linearly independent eigenfunctions, which is the required number.

\begin{figure}
	\centering
	\begin{overpic}[width = .45\textwidth]{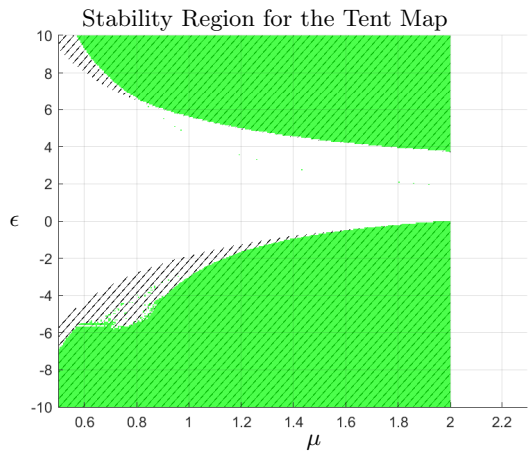}
	\put(54,2){$\mu$}
	\put(5,38){$\epsilon$}
	\put(17,71){Stability Region for the Tent Map}
	\end{overpic}
	\caption{The diagonal lines represent areas in which \eqref{eq:stability0}, \eqref{eq:instability1} and \eqref{eq:instability2} are satisfied for the tent map. The green region depicts $(\mu, \epsilon)$ values for which numerical simulations revealed non-stationary partial synchronization with different dynamics on each components.}
	\label{fig:StabilityRegionTentMap}
\end{figure}

We start by analyzing linear stability of the synchronized solution on $\cH_{c,t,\ell}$, which follows a similar pattern as for graphs as we just have to replace $\Delta_\cG$ by $\Delta_\cH$. First note, that for a synchronized solution to exist, we need to require $c=t$. Then, a necessary condition to retain at least partial synchronization ($x(i)=x(j)$ for some $i\neq j$) is stability in the direction of eigenfunctions, which are $+1$ on one vertex $-1$ on another vertex in the same component and $0$ everywhere else. As this is an eigenfunction corresponding to the eigenvalue $0$, \eqref{3} is equivalent to
\begin{equation}\label{eq:stability0}
	\mu_0<0.
\end{equation}
This is in clear contrast to the assumption $\mu_0>0$ for CMLs on graphs. In fact, on graphs the instability in direction of a spatially constant perturbation, which was caused by $\mu_0>0$, was necessary to have non-stationary dynamics of a synchronized solution. Given the condition $\mu_0<0$ on hyperflowers, the constant eigenfunction can no longer generate non-stationary dynamics. However, in contrast to $\Delta_\cG$, the hypergraph Laplace on the hyperflower has further eigenfunctions, which are constant on certain components of the hyperflower. By requiring instability of the synchronized solution with respect to perturbations in direction of these eigenfunctions, we may still hope to retain non-stationary dynamics of partially synchronized solutions. In other words, the eigenfunctions that are constant on each of the components and thus corresponding to positive eigenvalues are taking over the job of the constant eigenfunction corresponding to the eigenvalue $0$ on graphs. Instability in direction of the positive eigenvalue $\tilde \lambda = (c+t)/(c+t-1)$, which is responsible for differences between central and peripheral nodes, and $\hat \lambda = t/(c+t-1)$, that governs differences across the leafs, directly translates into the conditions
\begin{align}
	\label{eq:instability1}
	|\txte^{\mu_0}(1-\epsilon \tilde \lambda)|>1,\\
	\label{eq:instability2}
	|\txte^{\mu_0}(1-\epsilon \hat \lambda)|>1.
\end{align}
Even though one actually needs to find additional stability conditions around a partially synchronized solution, our numerical simulations reveal that the instability conditions around the completely synchronized solution do already provide great insight about the existence of non-stationary partially synchronized solutions. Especially, if $f$ is given by the tent map 
\begin{equation}\label{eq:tentmap}
	f(x) = \frac{\mu}{4}\left( 1-2\left| x-\frac 12\right|\right)
\end{equation}
this makes sense, as $f$ is piecewise linear and thus stability conditions derived from a linearization of $f(x)$ are to some extent independent of the particular state $x$. For the tent, the Lyapunov coefficient can explicitly be given by $\mu_0 = \ln(\mu/2)$. This allows us to further investigate which pairs $(\mu, \epsilon)$ fulfill the stability conditions \eqref{eq:stability0}, \eqref{eq:instability1} and \eqref{eq:instability2}. In particular, we marked areas in which the conditions are fulfilled by diagonal lines seen in Figure \ref{fig:StabilityRegionTentMap}. Further, a numerical integration of the system \eqref{5}, starting from a slight perturbation of a completely synchronized state, yields areas in which one has non-stationary partial synchronization with different dynamics in each of the components of the underlying hyperflower (see green regions in Figure \ref{fig:StabilityRegionTentMap}. As seen in Figure \ref{fig:SingleDynamics}, a closer look at the dynamics for parameter values in the green region shows chaotic dynamics on each component of the hyperflower. We observe several interesting phenomena. 

First, the results suggest that \eqref{eq:stability0} is sufficient for partial synchronization. Second, cluster synchronization of chaos only appears when the conditions \eqref{eq:stability0}, \eqref{eq:instability1} and \eqref{eq:instability2} are satisfied and third, chaotic dynamics can appear for values of $\mu<2$ for which the tent map alone exhibits no chaotic dynamics, but here a sufficiently positive or negative coupling induces chaos.

\begin{figure}
	\centering
	\begin{overpic}[width = .45\textwidth]{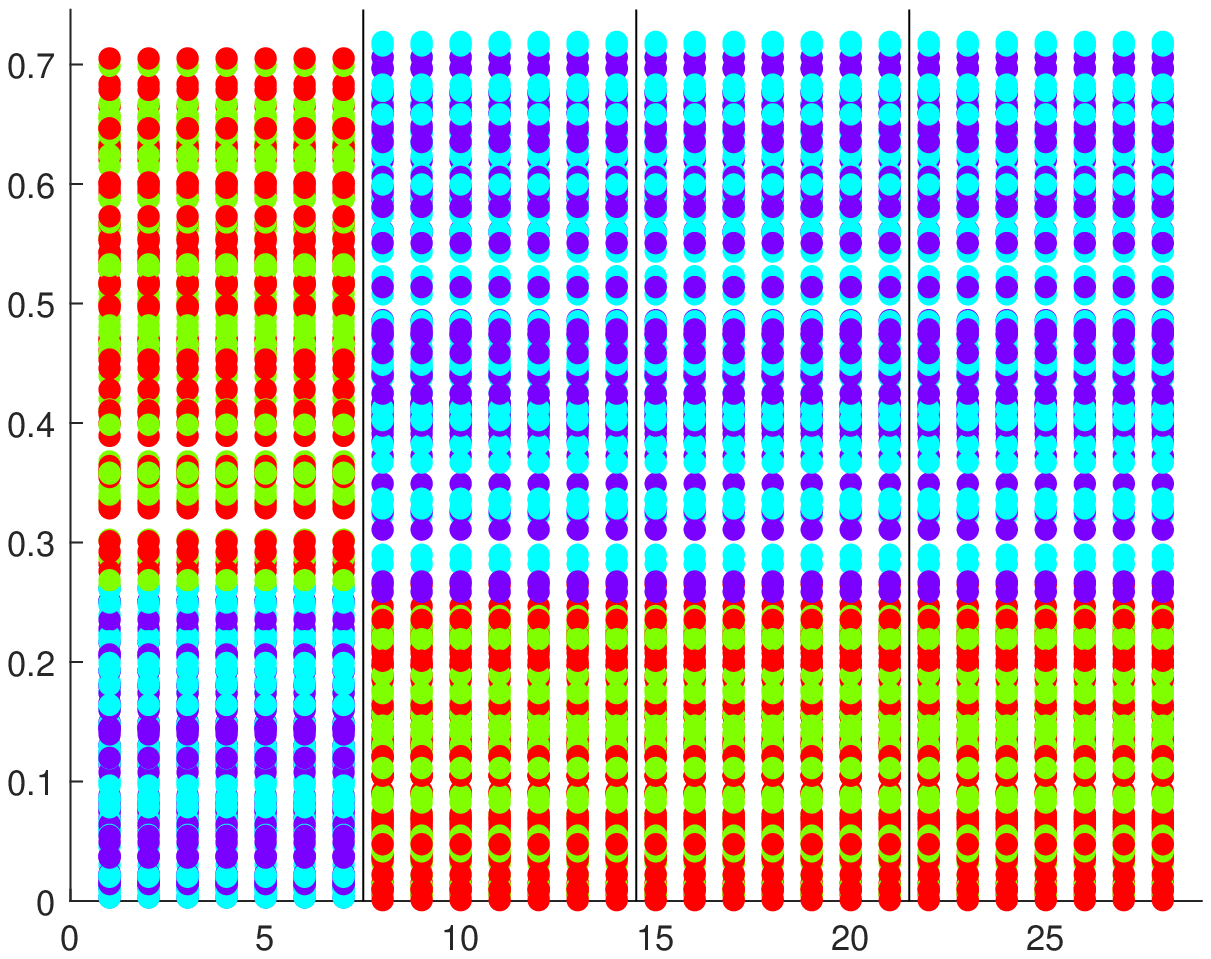}
		\put(3,35){\begin{turn}{90}$x_n(i)$\end{turn}}
		\put(18, 72){center}
		\put(36, 72){1st leaf}
		\put(54, 72){2nd leaf}
		\put(72, 72){3rd leaf}
		\put(48,1){Nodes $(i)$}		
	\end{overpic}
	\caption{Numerical Integration of \eqref{5} for $f(x)$ given by the tent map \eqref{eq:tentmap} on a hyperflower with $c=t=7$ and $\ell=3$ for $\mu = 1.8$ and $\epsilon = 3$. Plotted iterations are $5000< n \le 5200$. The values of $x_n(i)$ are alternately plotted in red, cyan, green and purple upon increasing $n$.}
	\label{fig:SingleDynamicsSyncLeafes}
\end{figure}

By neglecting the requirement of stability condition \eqref{eq:instability2}, i.e. allowing perturbations that are $-1$ on one leaf, $+1$ on another leaf and $0$ elsewhere to decay, we additionally observe parameter regions, in which all peripheral nodes synchronize among themselves and so do the central nodes but the two groups show different dynamics. For instance $(\mu, \epsilon) = (1.8, 3)$ satisfies \eqref{eq:stability0} and \eqref{eq:instability1} but not \eqref{eq:instability2}. The resulting dynamics can be seen in Figure \ref{fig:SingleDynamicsSyncLeafes}.

\begin{figure}
	\centering
	\begin{overpic}[width = .45\textwidth]{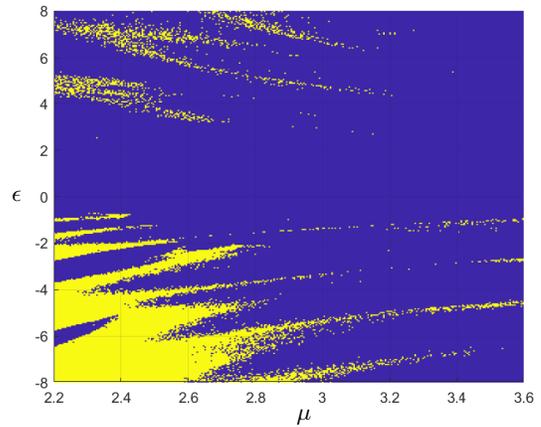}
	\put(53,2){$\mu$}
	\put(6,38){$\epsilon$}
	\end{overpic}
	\caption{Numerical simulations of \eqref{5} for $f(x)=\mu x(1-x)$ on a hyperflower with $c=10$, $\ell=5$ and $t=3$ revealed cluster synchronization of chaos in yellow regions. Doubly synchronized chaos occurs for all parameter values ($\mu, \epsilon)$ in the yellow regions.}
	\label{fig:HyperflowerRegion1}
\end{figure}

Even though our analytical derivations of stability conditions require assumptions about the hyperflower, numerical simulations can of course be performed for the cases not covered by our analytical derivations. Specifically, we consider simulations on a hyperflower with $c=10$, $\ell=5$ and $t=3$. For a given parameter pair $(\mu, \epsilon)$, we numerically infer synchronization of the central nodes if the standard deviation over $i = 1,\dots, c$ of $x_n(i)$ drops below a certain threshold ($\approx 10^{-5}$) as $n\to \infty$. Similarly, we infer chaos in the center of the hyperflower if the leading Lyapunov coefficient is positive on the central nodes. In the same way we deduce synchronization and chaotic behavior of nodes in the first leaf of the hyperflower. Based on those four criteria this allows us to classify the dynamical behavior for given parameter values and initial conditions. In particular, we say that the dynamics shows \emph{doubly synchronized chaos} if both of the leading Lyapunov coefficients for the two clusters are positive and the values of $x_n$ synchronize within the two clusters (but not necessarily across the clusters). Now, we conduct numerical simulations for different parameter values of $\mu$ and $\epsilon$ but with the same initial condition for each simulation and investigate for each parameter pair $(\mu, \epsilon)$ the occurrence of doubly synchronized chaos. The yellow regions in Figure \ref{fig:HyperflowerRegion1} depict such areas, whereas there is no doubly synchronized chaos in the blue region.

\begin{figure}
	\centering
	\begin{overpic}[width = .45\textwidth]{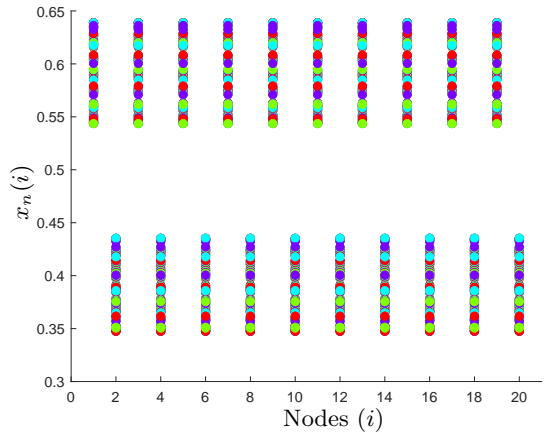}
		\put(3,35){\begin{turn}{90}$x_n(i)$\end{turn}}
		\put(48,1){Nodes $(i)$}		
	\end{overpic}
	\caption{Numerical Integration of \eqref{5} for $f(x) = \mu x(1-x)$ with $\mu = 2.868$ and $\epsilon = 6.04$ on a cyclic hypergrpah with $e=10$, $\ell = 6$, $m=1$ and $s=2$. Plotted iterations are $5000 < n \le 5200$. The values of $x_n(i)$ are alternately plotted in red, cyan, green and purple upon increasing $n$.}
	\label{fig:CyclicNoSymmetry}
\end{figure}

On hyperflowers, we have detected a variety of other patterns, including steady and periodic synchronization patters, as well as chaotic cluster patterns, where a single cluster chaotically forces clusters. Yet, above we have only shown the most complex interaction, where clusters are chaotically synchronized yet not correlated. Furthermore, we have considered less symmetric hypergraphs, e.g., the cyclic hypergraphs $\cZ_{e, \ell, m, s}$, which is a class defined by four parameters $e,\ell,m,s$. One can view $\cZ_{e, \ell, m, s}$ as a set of $es$ nodes, which are arranged in a circle. There are $e$ edges each encompassing $\ell$ neighbors. These edges are distributed uniformly around the circle such that if one edge starts at a node $i$ on the circle, the next edge starts at node that is $s$ nodes away from $i$. If one goes around the circle, the first $m$ nodes of each edge are specified as input nodes, whereas the remaining ones are output nodes.
While for some parameters, this class of hypergraphs has symmetries under permutation of nodes, it does not for others. If we consider, for example, the cyclic hypergraph with $e=10$ edges, $\ell = 6$, $m=1$ and $s=2$, there is no symmetric subgroup that leaves the hypergraph Laplace operator $\Delta_\cH$ invariant. Permuting two nodes would either cause edges to be spanned over non-neighboring nodes or edges not to start with nodes specified as input, both contradicting with a possible invariance of the hypergraph Laplacian. However, a numerical simulation starting from a completely synchronized initial condition with small perturbation, see Figure \ref{fig:CyclicNoSymmetry}, shows that both even and odd nodes form a cluster within which the dynamics synchronizes and shows chaotic behavior but there is no synchronization across the two clusters.

\section{Conclusion \& Outlook}

Although coupled map lattices (CMLs) have been a prototypical dynamical system studied on usual graphs for quite some time, so far no natural generalization to hypergraphs has been available. Here we provide this extension, which has been triggered by the requirement to model physical processes beyond pair-wise coupling. Classical CMLs show highly complex patterns due to the intertwining of Laplacian coupling and nonlinear iterated maps. Replacing the regular Laplacian by a hypergraph Laplacian led to new challenges. We used linearized stability analysis for synchronized states in combination with hypergraph spectral theory, and numerical methods, to detect robust regions of chaotic cluster synchronization for coupled hypergraph maps (CHMs). Chaotic cluster synchronization occurs in parameter space upon varying coupling strength and the main bifurcation parameter in the unimodal map at each node. We found key differences between Laplacian and hypergraph Laplacian coupling and detected also various other classes of periodic and quasi-periodic patterns. The results show the high complexity of CHMs. We expect that the generic nature of using a unimodal maps at each node and a generalization of the Laplacian should turn CHMs into an excellent universal model class for many concrete physical phenomena and to understand differences graph dynamics and hypergraph dynamics. For example, natural continuations of our work could aim to relate higher-order geometric structures induced by hypergraphs to generic dynamical phenomena, which is a line of research that has been very successful for studying CMLs and dynamics on networks more broadly. In addition, we anticipate that finding novel coarse-graining and mean-field methods will be needed to effectively carry out this research program for hypergraphs. CHMs are a natural starting point, since one can often build upon well-known and complete results for the individual nonlinear iterated maps at each node in low dimension as well as on studying the differences between more classical and hypergraph Laplace operators.\medskip

\begin{acknowledgments}
\textbf{Acknowledgments:} TB thanks the TUM Institute for Advanced Study (TUM-IAS) for support through a Hans Fischer Fellowship awarded to Chris Bick. TB also acknowledges support of the TUM TopMath elite study program. CK was supported a Lichtenberg Professorship of the VolkswagenStiftung. RM was supported by The Alan Turing Institute under the EPSRC grant EP/N510129/1.
\end{acknowledgments}

%\begin{references}
\bibliographystyle{unsrt}
\bibliography{Paper}
%\end{references}

\end{document}